\documentclass[11pt, Phys, Proceedings]{SciPost}

\hypersetup{
    colorlinks,
    linkcolor={red!50!black},
    citecolor={blue!50!black},
    urlcolor={blue!80!black}
}
\usepackage{mmacells}
\usepackage[varqu,varl,var0,scaled=0.97]{inconsolata}

\usepackage[bitstream-charter]{mathdesign}
\DeclareSymbolFont{usualmathcal}{OMS}{cmsy}{m}{n}
\DeclareSymbolFontAlphabet{\mathcal}{usualmathcal}
\newcommand{\munich}{Max-Planck-Institut f\"ur Physik, Werner-Heisenberg-Institut, 
80805 M\"unchen, Germany.}

\def\Mathematica{{{\sc Mathematica}}}

\def\Lotty{{{\sc Lotty}}}
\def\Fiesta{{{\sc Fiesta 4.2}}}
\def\SecDec{{{\sc SecDec 3.0}}}

\def\Cuba{{{\sc Cuba}}}

\usepackage{comment}

\begin{document}

\begin{center}{\Large \textbf{
Causal representation and numerical evaluation of multi-loop Feynman integrals\\
}}\end{center}

\begin{center}
William J. Torres~Bobadilla\textsuperscript{1$\star$}
\footnotetext{Parts of the work
presented in this talk have been carried out in collaboration with 
Paolo Benincasa, Germ\'an Rodrigo and Jonathan Ronca.}
\end{center}

\begin{center}
{\bf 1} \munich
* torres@mpp.mpg.de
\end{center}

\begin{center}
\today
\end{center}


\definecolor{palegray}{gray}{0.95}
\begin{center}
\colorbox{palegray}{
  \begin{tabular}{rr}
  \begin{minipage}{0.1\textwidth}
    \includegraphics[width=35mm]{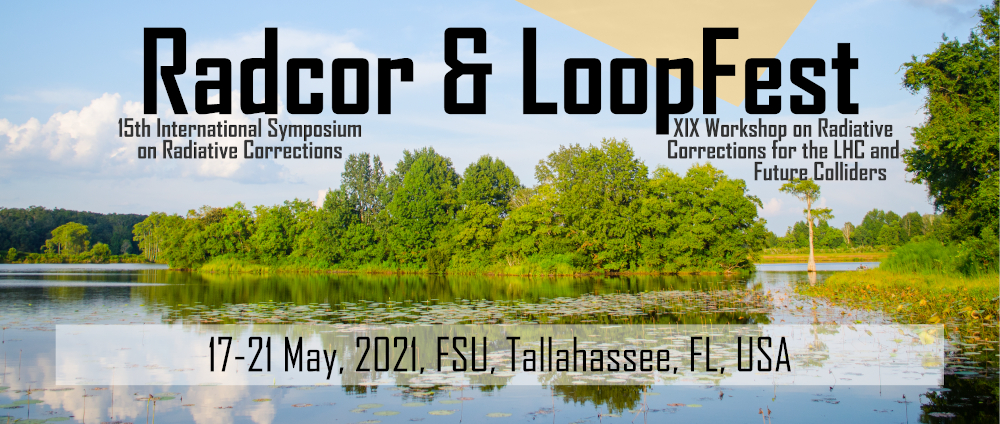}
  \end{minipage}
  &
  \begin{minipage}{0.85\textwidth}
    \begin{center}
    {\it 15th International Symposium on Radiative Corrections: \\Applications of Quantum Field Theory to Phenomenology,}\\
    {\it FSU, Tallahasse, FL, USA, 17-21 May 2021} \\
    \doi{10.21468/SciPostPhysProc.?}\\
    \end{center}
  \end{minipage}
\end{tabular}
}
\end{center}

\section*{Abstract}
{\bf
The loop-tree duality (LTD) has become a novelty
alternative to bootstrap the numerical evaluation of 
multi-loop scattering amplitudes. 
It has indeed been found that Feynman integrands,
after the application of LTD, display a representation 
containing only physical information, the so-called causal representation. 
In this talk, we discuss the all-loop causal representation
of multi-loop Feynman integrands, recently found in terms
of features that describe a loop topology, vertices and edges.
Likewise, in order to elucidate the numerical stability 
in LTD integrands, we present applications that 
involve numerical evaluations of 
two-loop planar and non-planar triangles 
with presence of several kinematic invariants.
}

\vspace{10pt}
\noindent\rule{\textwidth}{1pt}

\section{Introduction \& Motivation}
\label{sec:intro}

In this contribution, we report on the all-loop causal representation
of multi-loop scattering amplitudes, recently proposed in Ref.~\cite{TorresBobadilla:2021ivx}
and automated in the \Mathematica{} package \Lotty~\cite{Bobadilla:2021pvr}.
In this representation, that manifestly displays the causality of scattering amplitudes~\cite{aguilera2021manifestly}, the structure of Feynman integrands is cast
only by physical singularities, which, within our approach, are usually
referred to as causal propagators~\cite{Aguilera-Verdugo:2019kbz,Verdugo:2020kzh,Aguilera-Verdugo:2020kzc,Ramirez-Uribe:2020hes}.
Alternative approaches that aim to the generation of integrands displaying
only physical singularities have also 
been discussed in this conference~\cite{Capatti:2020ytd,Sborlini:2021owe,Ramirez-Uribe:2021ubp}. 

Interestingly, the causal representation, obtained as by-product of
the novel formulation of the loop-tree duality (LTD) formalism~\cite{Verdugo:2020kzh},
has allowed to have a complete understanding on the structure of multi-loop
integrands, regardless of the loop order. In particular, since LTD
relies on the direct application of the Cauchy residue theorem on
the energy component of the loop momenta~\cite{Catani:2008xa,Bierenbaum:2010cy,Bierenbaum:2012th,Buchta:2014dfa,Buchta:2015wna,
Jurado:2017xut,Driencourt-Mangin:2017gop,Aguilera-Verdugo:2019kbz,Verdugo:2020kzh,Plenter:2020lop,Driencourt-Mangin:2019aix,Driencourt-Mangin:2019yhu,
Tomboulis:2017rvd,Runkel:2019zbm,Runkel:2019yrs,Capatti:2019edf,Capatti:2019ypt,Capatti:2020ytd,Prisco:2020kyb,
deJesusAguilera-Verdugo:2021mvg}, 
causal integrands
are expressed in terms of on-shell energies, whose functional structure
turns out to be equivalent to the integrands obtained from the Old-Fashioned-Perturbation
theory (OFPT)~\cite{Weinberg:1995mt, Schwartz:2014sze}. 
However, by performing a term-by-term comparison between both approaches,
it turns out that the causal representation turns out to be more compact than OFPT,
which is due to redundant terms that cancel pairwise in the latter, 
as observed in Ref.~\cite{Capatti:2020ytd} of one-loop scalar integrands. 

In view of the structure of causal integrands obtained from LTD in 
Ref.~\cite{Aguilera-Verdugo:2020kzc}, we notice
several similarities when computing integrands whose 
loop topologies have the same number of 
\textit{vertices} but differ on the number of \textit{edges}. 
This observation promoted to a study in which causal integrands 
can be directly generated by means of the latter features 
and completely overcome the application of LTD. 
In order to carry out this study, we systematically consider 
all causal (or threshold) singularities of a given loop topology and organise 
them according to their compatibility. 
Likewise, as it was point in Ref.~\cite{TorresBobadilla:2021ivx},
the classification of loop topologies
by vertices and edges, as well as the generation 
of causal integrands, corresponds 
to a consequence of the well-known Steinmann relations~\cite{Steinmann:1960soa,Steinmann:1960sob, Araki:1961hb, Ruelle:1961rd, Stapp:1971hh, Lassalle:1974jm, Cahill:1973px,Cahill:1973qp}
and allowed us to provide a conjecture for the representation
of all-loop Feynman integrands.
Steinmann relations have been considered in recent works~\cite{Caron-Huot:2016owq,Caron-Huot:2019bsq,Benincasa:2020aoj,Caron-Huot:2020bkp,Bourjaily:2020wvq}
that aim at bootstrapping analytic structure of 
multi-loop scattering amplitudes. 

Furthermore, inspired by the recent developments of LTD to numerically evaluate
multi-loop Feynman integrals and keep understanding 
mathematical properties to causal integrands, 
we provided the publicly \Mathematica{} package \Lotty{}~\cite{Bobadilla:2021pvr} that 
illustrates and automates the generation of dual and causal integrands,
allowing, in this way, to provide the scientific community with 
alternative strategies to overcome the obstacles in the calculation 
of multi-loop scattering amplitudes~\cite{Gnendiger:2017pys,Heinrich:2020ybq,TorresBobadilla:2020ekr}. 

\par\bigskip
This contribution is organised as follows. In Sec.~\ref{sec:causal} 
we briefly remark the main features of the 
causal representation of multi-loop scattering amplitudes.
Then, in Sec.~\ref{sec:num_eval}, we recall 
numerical stability and numerical integration of 
causal integrands.
Then, in Sec.~\ref{sec:discussion}, we draw a summary of the talk
and discuss future directions.

\section{All-loop causal representation of Feynman integrands}
\label{sec:causal}

\begin{figure}[t]
\centering
\includegraphics[scale=1]{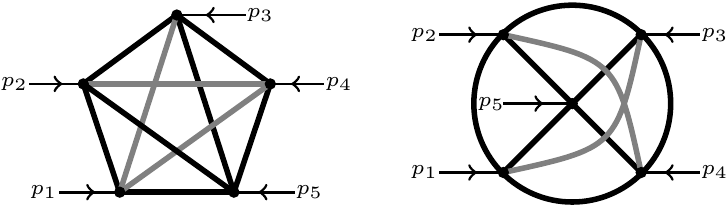}
\caption{Multi-loop topology with five vertices and ten internal lines.}
\label{eq:5cusps}
\end{figure}

In order to explicitly illustrate the causal representation of multi-loop
Feynman integrands, let us consider, for the sake of simplicity,
the most general loop topology built from five vertices and all 
possible connections among edges (or internal lines).
A pictorial representation of this six-loop topology is depicted in Fig.~\ref{eq:5cusps},
in which grey lines do not cross among themselves. 
Hence, taking into account this topology, one can consider a 
five-point six-loop Feynman integral with ten internal lines, 
\begin{align}
\mathcal{A}_{5}^{\left(6\right)}\left(1,2,\hdots,10\right) & =\int_{\ell_{1},\hdots,\ell_{6}}
\mathcal{N}\left(\left\{ \ell_{i}\right\} _{6},\left\{ p_{j}\right\} _{5}\right)\times G_{F}\left(1,\hdots,10\right)\,,
\label{eq:myamp}
\end{align}
where
$\int_{\ell_{s}}\equiv-\imath\mu^{4-d}\int d^{d}\ell_{s}/\left(2\pi\right)^{d}$ and 
Feynman propagators $G_F$ are compactly represented as follows, 
\begin{align}
G_{F}\left(1,\hdots,r\right) & =\prod_{i=1}^{r}\left(G_{F}\left(q_{i}\right)\right)^{\alpha_{i}}\,,
\end{align}
where $\alpha_{i}$ are powers of the propagators and,
motivated by LTD, one explicitly pulls out the dependence on 
the energy components of the loop momenta in each internal line, 
\begin{align}
G_{F}\left(q_{i}\right) & =\frac{1}{q_{i}^{2}-m_{i}^{2}+\imath0}=\frac{1}{\left(q_{i,0}+q_{i,0}^{\left(+\right)}\right)\left(q_{i,0}-q_{i,0}^{\left(+\right)}\right)}\,.
\end{align}
The latter corresponds to the usual Feynman propagator of a one single particle, with $m_{i}$
its mass, $+\imath0$ the infinitesimal imaginary Feynman prescription
and, 
\begin{align}
q_{i,0}^{\left(+\right)} & =\sqrt{\boldsymbol{q}_{i}^{2}+m_{i}^{2}-\imath0}\,,
\end{align}
the on-shell energy of the loop momentum $q_{i}$ written in terms
of the spatial components $\boldsymbol{q}_{i}$. 

\par\bigskip 
Then, to describe the loop topology of Fig.~\ref{eq:5cusps}, 
we work with the propagators, 
\begin{align}
q_{i}=&\begin{cases}
\ell_{i},\, & i=1,\hdots,6\\
\ell_{1}-\ell_{4}-\ell_{5}-p_{1}, & i=7\\
\ell_{2}-\ell_{1}-\ell_{6}-p_{2}, & i=8\\
\ell_{3}-\ell_{2}+\ell_{5}-p_{3}, & i=9\\
\ell_{4}-\ell_{3}+\ell_{6}-p_{4} & i=10
\end{cases}
\quad \,,
\end{align}
in which, after applying LTD, one recognises the set of causal propagators, 
\begin{align}
 & \lambda_{1}^{\pm}=q_{\left(1,4,5,7\right),0}^{\left(+\right)}\pm p_{1,0}\,, &  & \lambda_{12}^{\pm}=q_{\left(2,4,5,6,7,8\right),0}^{\left(+\right)}\pm p_{12,0}\,, &  & \lambda_{24}^{\pm}=q_{\left(1,2,3,4,8,10\right),0}^{\left(+\right)}\pm p_{24,0}\,,\nonumber \\
 & \lambda_{2}^{\pm}=q_{\left(1,2,6,8\right),0}^{\left(+\right)}\pm p_{2,0}\,, &  & \lambda_{13}^{\pm}=q_{\left(1,2,3,4,7,9\right),0}^{\left(+\right)}\pm p_{13,0}\,, &  & \lambda_{35}^{\pm}=q_{\left(2,3,5,7,8,10\right),0}^{\left(+\right)}\pm p_{35,0}\,,\nonumber \\
 & \lambda_{3}^{\pm}=q_{\left(2,3,5,9\right),0}^{\left(+\right)}\pm p_{3,0}\,, &  & \lambda_{23}^{\pm}=q_{\left(1,3,5,6,8,9\right),0}^{\left(+\right)}\pm p_{23,0}\,, &  & \lambda_{34}^{\pm}=q_{\left(2,4,5,6,9,10\right),0}^{\left(+\right)}\pm p_{34,0}\,,\nonumber \\
 & \lambda_{4}^{\pm}=q_{\left(3,4,6,10\right),0}^{\left(+\right)}\pm p_{4,0}\,, &  & \lambda_{45}^{\pm}=q_{\left(3,4,6,7,8,9\right),0}^{\left(+\right)}\pm p_{45,0}\,, &  & \lambda_{25}^{\pm}=q_{\left(1,2,6,7,9,10\right),0}^{\left(+\right)}\pm p_{25,0}\,,\nonumber \\
 & \lambda_{5}^{\pm}=q_{\left(7,8,9,10\right),0}^{\left(+\right)}\pm p_{5,0}\,, &  & \lambda_{14}^{\pm}=q_{\left(1,3,5,6,7,10\right),0}^{\left(+\right)}\pm p_{14,0}\,, &  & \lambda_{15}^{\pm}=q_{\left(1,4,5,8,9,10\right),0}^{\left(+\right)}\pm p_{15,0}\,.\label{eq:lambn3mlt}
\end{align}

\begin{figure}[t]
\centering
\includegraphics[scale=1]{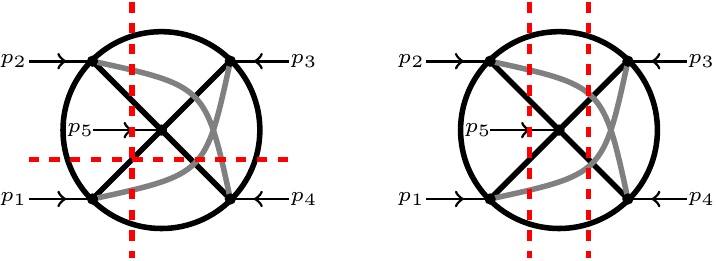}
\caption{Loop topologies with (\textit{left}:) two overlapped and 
(\textit{right}:) two entangled thresholds.}
\label{fig:overlap}
\end{figure}

The causal representation of a topology built from five vertices
is expected to admit the decomposition in terms of the following summands, 
\begin{align}
\frac{1}{\lambda_{i}^{h_{i}}\lambda_{j}^{h_{j}}\lambda_{k}^{h_{k}}\lambda_{l}^{h_{l}}}\,,
\end{align}
where $i,j,k,l\in\{1,2,3,4,5,12,13,14,15,23,24,25,34,35\}$
and $h_{\{i,j,k,l\}}\in\{+,-\}$. 
Nevertheless, not all products of $\lambda_i$'s in~\eqref{eq:lambn3mlt} will appear
in the representation, since there will be products of two $\lambda_{i}^{\pm}$'s, 
that cannot appear because
of the incompatible alignment of internal and external momenta. 
To illustrate this subtlety, let us consider 
a possible entanglement between $\lambda_{12}^{\pm}$, $\lambda_{14}^{\pm}$
and $\lambda_{12}^{\pm}$, $\lambda_{34}^{\pm}$. 
In Fig.~\ref{fig:overlap} (left), we note that the causal thresholds 
$\lambda_{12}^{\pm}$, $\lambda_{14}^{\pm}$ cannot be considered together
when providing a candidate for the causal representation,
whereas, $\lambda_{12}^{\pm}$, $\lambda_{34}^{\pm}$,
in Fig.~\ref{fig:overlap} (right), appear
with combinations of other $\lambda_{i}^{\pm}$'s.
The very same analysis can be applied to tuples of length three and four.

Interestingly, the above-mentioned features of overlapped and entangled causal thresholds
may be related to
Steinmann relations, 
since they state that double discontinuities 
in overlapping cuts vanish,
\begin{align}
&\text{Disc}_{s_{I}}\left(\text{Disc}_{s_{J}}\mathcal{A}_{N}^{\left(L\right)}\right)=0\,,\quad\text{where }
\begin{cases}
I\not\subset J\\
J\not\subset I
\end{cases}
\,,
\label{eq:steinmann_rel}
\end{align}
where  
$\text{Disc}_{s_I}\mathcal{A}_{N}^{\left(L\right)}$ represents the discontinuity of 
the scattering amplitude (or Feynman integral) of Eq.~\eqref{eq:myamp} 
at the kinematic invariant 
$s_{I}=\left(\sum_{i\in I}p_{i}\right)^{2}$ generated by the external momenta
that belong to the set $I$.
A naive interpretation of the Steinmann relations within our framework
may be given by promoting the kinematic invariants $s_I$ to the 
sum of energies of external momenta present in the causal propagators. 

Hence, by constructing an Ansatz that accounts for the Steinmann relations 
and reconstructing the analytic expressions for the causal integrand 
by means of finite fields arithmetic~\cite{vonManteuffel:2014ixa,Peraro:2016wsq,Klappert:2019emp,Peraro:2019svx}, we find for the six-loop five-point scalar integral,
\begin{align}
&\mathcal{A}_{5}^{\left(6\right)}=\int_{\ell_{1},\hdots,\ell_{6}}\frac{1}{x_{10}}
 \sum_{\substack{
i=1\\
j=i+1}}^{5}
\ \sum_{\substack{k=1\\
l=k+1\\
k,l\ne i,j
}
}^{5}
L_{ij}^{+}\,L_{kl}^{-}\,,
&& 
L_{ij}^{\pm}=\frac{1}{\lambda_{ij}^{\pm}}\left(\frac{1}{\lambda_{i}^{\pm}}+\frac{1}{\lambda_{j}^{\pm}}\right)\,,
 \label{eq:n3mltcausal}
\end{align}
with $x_{n}=\prod_{i=1}^{n}2\,q_{i,0}^{\left(+\right)}$. 

Notice that, in view of this very compact structure of 
the causal representation of a six-loop integrand with five vertices, 
we proposed in Ref.~\cite{TorresBobadilla:2021ivx} an algorithm to generate causal 
representation of topologies constructed from $k+2$ cusps.
In fact, due to symmetry in the structure of  topologies when all possible connections, edges, 
between vertices are considered, we conjectured a close formula when 
all possible edges are taken into account,
\begin{align}
\mathcal{F}_{L+k} = & \sum_{\substack{i_{1}\ll i_{N_{i}}\\
j_{1}\ll j_{N_{j}}
}
}^{k+2}\Omega_{\vec{i}}^{\vec{j}}\ L_{i_{1}i_{2}\hdots i_{N_{i}}}^{+}L_{j_{1}j_{2}\hdots j_{N_{j}}}^{-}\,,
\label{eq:allcausal}
\end{align}
with, 
\begin{align}
\Omega_{\vec{i}}^{\vec{j}}&=
\begin{cases}
1 & \text{If }\vec{i}\cap\vec{j}=\emptyset\\
0 & \text{otherwise}
\end{cases}\,,
\end{align}
where $\vec{i}=\{i_1,i_2,\hdots,i_{N_i}\}$, $\vec{j}=\{j_1,j_2,\hdots,j_{N_j}\}$,
 $i_{1}\ll i_{N_{i}}$ is the lexicographic ordering $i_{1}<i_{2}<\cdots<i_{N_{i}}$,
 $N_{i}=\left[k/2\right]+1$ and $N_{j}=k-\left[k/2\right]$. 
The functions $L_{ijk\hdots}^{\pm}$ contain the causal information of the integrand,
\begin{align}
 & L_{i_{1}i_{2}\hdots i_{N}}^{\pm}=\frac{1}{\lambda_{i_{1}i_{2}\hdots i_{N}}^{\pm}}\sum_{\substack{j_{1}\ll j_{N-1}\\
\vec{j}\subset\vec{i}
}
}L_{j_{1}j_{2}\hdots j_{N-1}}^{\pm}\,, &  & L_{i_{1}}^{\pm}=\frac{1}{\lambda_{i_{1}}^{\pm}}\,.
\label{eq:allLs}
\end{align}
The close formula~\eqref{eq:allcausal}, together with the definitions of $L^\pm$, 
have been implemented in the built-in routines of the \Mathematica{} package
\Lotty{} (see Sec.~\ref{sec:num_eval}).

On top of decomposition~\eqref{eq:allcausal}, we would like to 
remark that an extensive study of various representations of 
scattering amplitudes, including the ones presented in this talk
(LTD and causal), are currently 
under consideration~\cite{Benincasa:2021xyz},
where we elucidate a clear pattern to go from one representation
to another along the lines of Ref.~\cite{Arkani-Hamed:2017fdk}. 

\section{Numerical evaluation of Feynman integrals}
\label{sec:num_eval}
\begin{figure}[t!]
\includegraphics[scale=0.75]{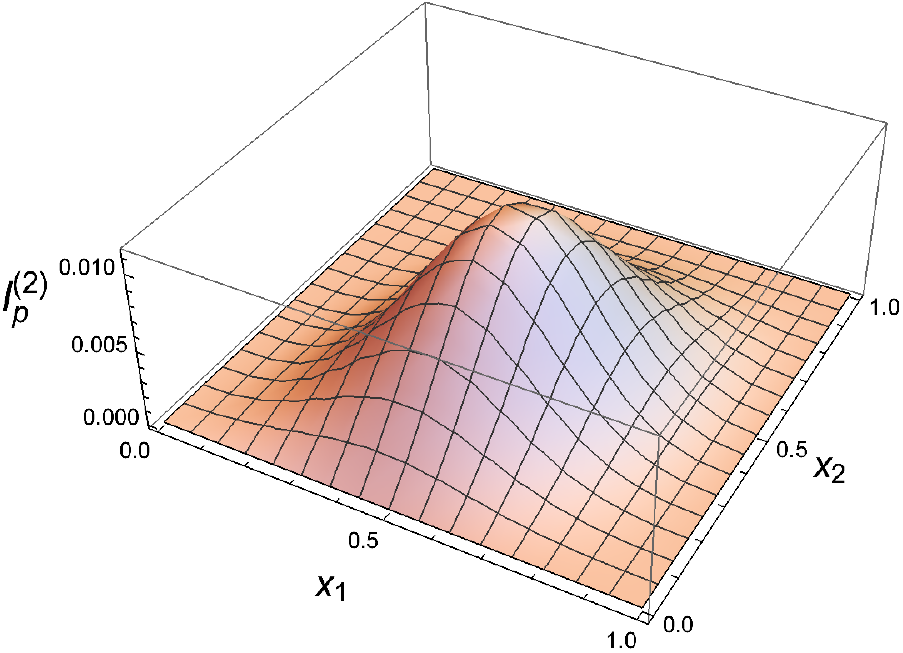}\quad
\includegraphics[scale=0.75]{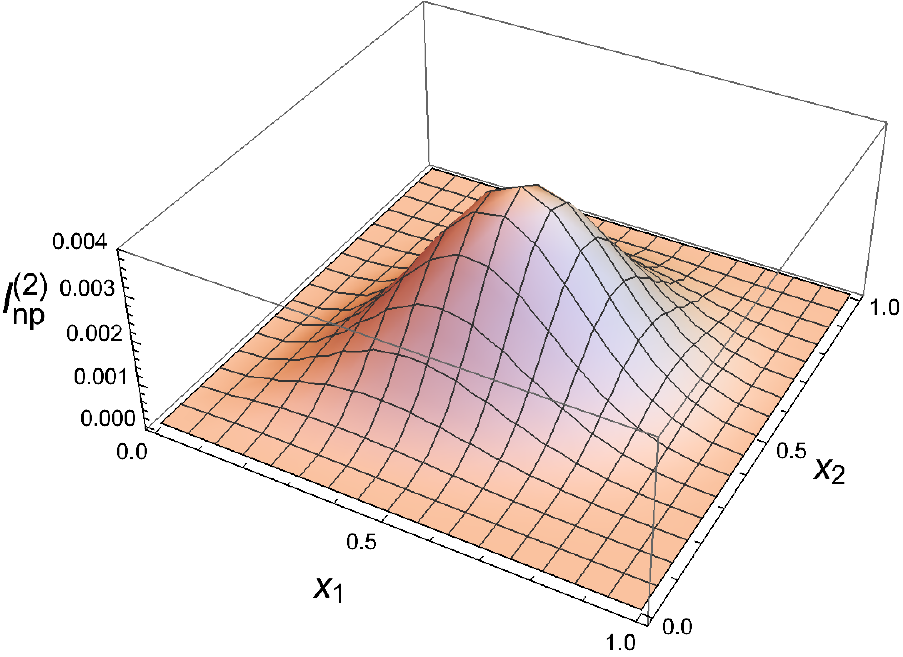}
\caption{Numerical stability of causal integrands of 
two-loop planar (left) and non-planar (right) scalar triangles
as a function of $x_1$ and $x_2$ with fixed values for the angles $\theta_{ij}$. 
}
\label{fig:3d_tris}
\end{figure}
In Refs.~\cite{Bobadilla:2021pvr}, we provided the \Mathematica{} package \Lotty{}
that automates dual and causal representation of LTD along the lines
of Refs.~\cite{Verdugo:2020kzh} and~\cite{TorresBobadilla:2021ivx}, respectively. 
This package is publicly available and can be downloaded from the 
Bitbucket repository:
\begin{equation*}
  \textrm{\href{https://bitbucket.org/wjtorresb/lotty/src/master/}{\color{blue}git clone https://wjtorresb@bitbucket.org/wjtorresb/lotty.git}}
\,.
\end{equation*}
The main motivation to provide this automation was to give the scientific community
with an illustrative and pedestrian overview of  LTD by means 
of a \Mathematica{} package. 
In effect, to recreate the advantage of employing the causal representation of 
multi-loop scattering amplitudes, 
we provide the evaluation of 
planar ($I_{\text{p}}^{(2)}$) and non-planar ($I_{\text{np}}^{(2)}$)
three-point integrals at two loops.

Furthermore, because of the causal representation of multi-loop 
scalar integrands, conjectured in Ref.~\cite{TorresBobadilla:2021ivx} and 
briefly summarised in Sec.~\ref{sec:causal}, 
one can simply  make use of the built-in \Lotty{} routine \verb"AllCausal" 
and get the most compact representation of the latter 
in terms of physical singularities, 
{ \small
\begin{mmaCell}[moredefined={AllCausal}]{Input}
AllCausal[5]
\end{mmaCell}
}\vspace{-3ex}
{ \small
\begin{mmaCell}{Output}
  (L[-][\{3,4\}] + L[-][\{3,5\}] + L[-][\{4,5\}]) L[+][\{1,2\}]
+ (L[-][\{2,4\}] + L[-][\{2,5\}] + L[-][\{4,5\}]) L[+][\{1,3\}]
+ (L[-][\{2,3\}] + L[-][\{2,5\}] + L[-][\{3,5\}]) L[+][\{1,4\}]
+ (L[-][\{2,3\}] + L[-][\{2,4\}] + L[-][\{3,4\}]) L[+][\{1,5\}]
+ (L[-][\{1,4\}] + L[-][\{1,5\}] + L[-][\{4,5\}]) L[+][\{2,3\}]
+ (L[-][\{1,3\}] + L[-][\{1,5\}] + L[-][\{3,5\}]) L[+][\{2,4\}]
+ (L[-][\{1,3\}] + L[-][\{1,4\}] + L[-][\{3,4\}]) L[+][\{2,5\}]
+ (L[-][\{1,2\}] + L[-][\{1,5\}] + L[-][\{2,5\}]) L[+][\{3,4\}]
+ (L[-][\{1,2\}] + L[-][\{1,4\}] + L[-][\{2,4\}]) L[+][\{3,5\}]
+ (L[-][\{1,2\}] + L[-][\{1,3\}] + L[-][\{2,3\}]) L[+][\{4,5\}]
\end{mmaCell}
}
\noindent
where, as mentioned in Sec.~\ref{sec:causal}, each $L[\pm][\hdots]$
is expressed in terms of physical singularities only. 
Besides, to explicitly represent causal representations 
for planar and non-planar triangles in terms of on-shell energies, 
one can yet make use of the additional \Lotty{} routines: 
\verb"RefineCausal" and \verb"Lamb2qij". 

Let us remark that within our automation in \Mathematica{}
we are not intending to provide a full numerical integration
of multi-loop Feynman integrals. 
Moreover, we can profit from the latter and naively integrate our
causal integrands through the built-in function of \Lotty{}
for the change of variables, e.g. a two-loop integrand in $d=4$ dimensions:
{\small
\begin{mmaCell}[moredefined={LoopMom,dim}]{Input}
dim = 4;		  (* set the dimension *)
LoopMom = \{l1, l2\};	(* list of loop momenta *)
\end{mmaCell}
}\vspace{-3ex}
{\small
\begin{mmaCell}[moredefined={LoopMom,dim,LoopToSC,meassure,intvars}]{Input}
LoopToSC[LoopMom, dim]
meassure[LoopMom, dim]
intvars[LoopMom, dim, "MaxValueR"->max]
\end{mmaCell}
}\vspace{-3ex}
{\small
\begin{mmaCell}{Output}
\{l1 -> \{r1 Cos[\(\theta\)11], r1 Cos[\(\theta\)12] Sin[\(\theta\)11],\
r1 Sin[\(\theta\)11] Sin[\(\theta\)12] \}, 
 l2 -> \{r2 Cos[\(\theta\)21], r2 Cos[\(\theta\)22] Sin[\(\theta\)21],\ 
r2 Sin[\(\theta\)21] Sin[\(\theta\)22] \} \}
\end{mmaCell}
}\vspace{-3ex}
{\small
\begin{mmaCell}{Output}
\mmaSup{r1}{2} \mmaSup{r2}{2} Sin[\(\theta\)11] Sin[\(\theta\)21]
\end{mmaCell}
}\vspace{-3ex}
{\small
\begin{mmaCell}{Output}
\{\{r1,0,max\}, \{r2,0,max\}, \{\(\theta\)11,0,\(\pi\)\}, \{\(\theta\)12,0,\
2\(\pi\)\}, \{\(\theta\)21,0,\(\pi\)\}, \{\(\theta\)22,0,2\(\pi\)\}\}   
\end{mmaCell}
}\vspace{-1ex}
\noindent
where, by default in the integration limits of each radius, 
\texttt{"MaxValueR"->$10^5$}.
However, to avoid considering the integration domain $[0,\infty)$,
one can map $[0,\infty)$ to the segment $(0,1]$, 
with the change of variables, 
$r_i = (1-x_i)/x_i$\,.
Additionally, expressions for on-shell energies ($q_{i,0}^{(+)}$)
can be also generated by \Lotty{}, taking into account 
the momentum components of the external kinematics. 
For instance, in the three-point loop topology, one may choose,
without loss of generality, the center-of-mass frame (COMF),
\begin{align}
& p_1^\mu=E_\text{cm}/2\{1,1,0,0\} && p_2^\mu=E_\text{cm}/2\{1,-1,0,0\}\,,
\end{align}
where $E_{\text{cm}}$ is the energy of COMF. 
In the two-loop planar triangle, this amounts to, 
{\small
\begin{mmaCell}{Output}
\{\mmaSubSup{q[1]}{0}{(+)} ->  \mmaSqrt{\mmaSup{r1}{2} + \mmaSup{m[1]}{2}}, 
 \mmaSubSup{q[2]}{0}{(+)} -> \mmaFrac{1}{2}\mmaSqrt{\mmaSup{Ecm}{2} + 4 Ecm r1 Cos[\(\theta\)11] + 4 (\mmaSup{r1}{2} + \mmaSup{m[2]}{2})}, 
 \mmaSubSup{q[3]}{0}{(+)} ->  \mmaSqrt{\mmaSup{r1}{2} + \mmaSup{m[3]}{2}}, 
 \mmaSubSup{q[4]}{0}{(+)} ->  \mmaSqrt{\mmaSup{r2}{2} + \mmaSup{m[4]}{2}}, 
 \mmaSubSup{q[5]}{0}{(+)} ->  \mmaSqrt{\mmaSup{r2}{2} + \mmaSup{m[5]}{2}}, 
 \mmaSubSup{q[6]}{0}{(+)} ->  \mmaSqrt{\mmaSup{r1}{2}+\mmaSup{r2}{2}+\mmaSup{m[6]}{2}+2r1r2 (Cos[\(\theta\)11]Cos[\(\theta\)21] + Cos[\(\theta\)12-\(\theta\)22]Sin[\(\theta\)11]Sin[\(\theta\)21])} \}
\end{mmaCell}
}\noindent
The numerical stability of causal integrands for 
planar and non-planar two-loop triangles in terms of 
the compactified variable $x_i$
and angles $\theta_{ij}$ is displayed in Fig.~\ref{fig:3d_tris}.
\begin{table}
\centering
\begin{tabular}{|c|c|c|c|c|}
\hline
\multicolumn{1}{|c}{}& \multicolumn{2}{|c|}{\textbf{Planar triangle}} & \multicolumn{2}{c|}{\textbf{Non-planar triangle}} \\
\hline 
$\frac{s}{m^{2}}$ & LTD ($10^{-6}$) & \SecDec{} ($10^{-6}$) & LTD ($10^{-6}$) & \SecDec{} ($10^{-6}$)\tabularnewline
\hline 
\hline 
$-\frac{1}{4}$ & 9.48(5) & 9.4647(9) & 4.461(3)  & 4.4606(4)\tabularnewline
\hline 
$-1$ & 8.10(5) & 8.0885(8) & 4.101(3)  & 4.1012(4)\tabularnewline
\hline 
$-\frac{9}{4}$ & 6.49(3) & 6.4760(6) & 3.627(5)  & 3.6276(3)\tabularnewline
\hline 
$-4$ & 5.02(2) & 5.0188(5) & 3.15(5)  & 3.1334(3)\tabularnewline
\hline 
$+\frac{1}{4}$ & 10.68(6) & 10.651(1) & 4.743(3)  & 4.7436(4)\tabularnewline
\hline 
$1$ & 13.11(8) & 13.070(1) & 5.259(3)  & 5.2590(5)\tabularnewline
\hline 
$+\frac{9}{4}$ & 20.81(1) & 20.748(2) & 6.533(3)  & 6.5331(6)\tabularnewline
\hline 
$+\frac{25}{16}$ & 15.74(9) & 15.700(1) & 5.748(3)  & 5.7474(6)\tabularnewline
\hline 
\end{tabular}
\caption{Numerical integration with \texttt{NIntegrate} 
of planar and non-planar two-loop triangles 
in the causal representation.
The values of integrations are compared with \SecDec. 
}
\label{tab:numint}
\end{table}
Hence, one can naively make use of the built-in \Mathematica{} routine \verb"NIntegrate"
to integrate our expressions. 
A few representative results are collected in Table~\ref{tab:numint},
where we emphasise that no optimisation or educated use of the latter routine
is carried out. 
Likewise, to compare our numerical integrations, we make use of 
publicly automated codes that evaluates the latter
by means of the sector decomposition algorithm~\cite{Hepp:1966eg,Roth:1996pd,Binoth:2000ps,Heinrich:2008si}, 
\SecDec~\cite{Borowka:2015mxa} 
and \Fiesta~\cite{Smirnov:2015mct}.

In view of the interesting compact structure obtained for multi-loop 
causal integrands, it is worth dedicating further studies to numerical 
integrations. The natural extension is having a stable integrator
that systematically allows to improve precision in evaluations. 
To this end, one can rely on publicly available implementations 
for multi-dimensional numerical integrations, e.g. \Cuba{} library~\cite{Hahn:2004fe,Hahn:2016ktb},
and interface it with \Lotty{} to, thus, keep control 
of the numerical evaluation from the symbolic side~\cite{Rodrigo:2021xyz}. 

\section{Discussion}
\label{sec:discussion}
The loop-tree duality (LTD) formulation has become a novel alternative 
approach to numerically
evaluate multi-loop scattering amplitudes, which 
in the current era is one of the main obstacles to go further 
in theoretical predictions. 
However, dual integrands, or expressions obtained from the direct application
of LTD on multi-loop Feynman integrands, are not the ultimate representation,
since it yet contains so-called spurious contributions but they can yet be 
employed to understand LTD decompositions in scattering amplitudes,
regardless of the loop order~\cite{Verdugo:2020kzh}. 

In this talk, we introduced the concept of causal representation 
of multi-loop scattering amplitudes, which in a nutshell corresponds 
to having an integrand displaying only physical singularities. 
Remarkably, the Old-Fashioned-Perturbation theory (OFPT) aims at 
displaying a very same pattern for Feynman integrands.
However, these OFPT integrands turn out to be explicitly classified according to 
number of vertices and not accounting for number of edges (or internal lines). 
In the causal representation, generated from LTD, 
the situation is slightly different, the structure of causal integrands 
accounts for both vertices and edges. 
In effect, as elucidated in Ref.~\cite{TorresBobadilla:2021ivx}, 
it is possible to generate a representation for all-loop scalar integrands,
by accounting for the features of the loop topology,
given by vertices and edges. 

Furthermore, we remarked that the all-loop causal representation 
for multi-loop Feynman integrands 
has been conjectured by keeping into account the Steinmann relations.
Moreover, to give support to this conjecture, we provided a \Mathematica{}
implementation of the latter by means of the automated package \Lotty{}~\cite{Bobadilla:2021pvr}, 
in which a direct comparison between LTD and causal integrands
can straightforwardly be performed. 

Different representations of scattering amplitudes are currently appearing,
the main question to address is: what is then the best representation? 
Perhaps, there is no a unique answer and, on the contrary, 
one could think of asking further and being more specific.
For instance, what would we like to extract from 
a given representation? 
In this case, one could put on the table all available representations and 
extract the most convenient one for the desirable study~\cite{Benincasa:2021xyz}.

In the spirit of profiting from the compact structure of 
the causal representation of multi-loop Feynman integrands, 
we observed a smooth behaviour in their numerical evaluation.
Therefore, to integrate the latter,  this is a very important insight 
since one can only focus on the treatment of physical singularities. 
We illustrated that by simply using the built-in \Mathematica{} routines 
with no optimisation a preliminary numerical integration can yet be carried out. 
However, it is worth investigating further the interplay between 
analytic expressions provided by \Lotty{} and the various 
implementations that aim at having full control 
on multi-dimensional numerical integration~\cite{Rodrigo:2021xyz}.

\section*{Acknowledgements}
The author wishes to thank 
Nima Arkani-Hamed, 
Andrea Banfi, 
Martin Beneke,
Johannes Henn,
Jonas Lindert, 
Sebastian Mizera, 
and 
Tiziano Peraro
for enlightening discussions. 




\bibliographystyle{SciPost_bibstyle} 
\bibliography{refs}

\nolinenumbers

\end{document}